\documentclass[11pt]{article}
\usepackage[margin=1in]{geometry}
\usepackage[colorlinks=true, pdfborder={0 0 0}, citecolor=blue, linkcolor=blue]{hyperref}
\usepackage{graphicx}
\usepackage{authblk}
\renewcommand\Authfont{\normalsize}
\renewcommand\Affilfont{\itshape\footnotesize}
\usepackage{siunitx}
\usepackage{chemformula}
\usepackage{url}
\usepackage[labelfont=bf]{caption}
\usepackage{setspace}
\usepackage{upgreek}
\usepackage{helvet}
\usepackage[numbers,square,sort&compress]{natbib}
\usepackage[resetlabels]{multibib}
\newcites{Sup}{References}

\title{{\Large \textbf{High performance integrated graphene electro-optic modulator at cryogenic temperature}}}

\author[1]{Brian S. Lee}
\author[2]{Bumho Kim}
\author[3]{Alexandre P. Freitas}
\author[1]{Aseema Mohanty}
\author[2]{Yibo Zhu}
\author[1]{Gaurang R. Bhatt}
\author[2]{James Hone}
\author[1,*]{Michal Lipson}
\affil[1]{Department of Electrical Engineering, Columbia University, New York, NY 10027, USA}
\affil[2]{Department of Mechanical Engineering, Columbia University, New York, NY 10027, USA}
\affil[3]{School of Electrical and Computer Engineering, University of Campinas, Campinas-SP, 13083-970, Brazil}
\affil[*]{Corresponding author: ml3745@columbia.edu}
\date{}

\begin{document}
	\maketitle
	\vspace{-3em}
	\doublespacing
	
	\section*{Abstract}
	\textbf{High performance integrated electro-optic modulators operating at low temperature are critical for optical interconnects in cryogenic applications. Existing integrated modulators, however, suffer from reduced modulation efficiency or bandwidth at low temperatures because they rely on tuning mechanisms that degrade with decreasing temperature. Graphene modulators are a promising alternative, since graphene's intrinsic carrier mobility increases at low temperature. Here we demonstrate an integrated graphene-based electro-optic modulator whose 14.7 GHz bandwidth at 4.9 K exceeds the room-temperature bandwidth of 12.6 GHz. The bandwidth of the modulator is limited only by high contact resistance, and its intrinsic RC-limited bandwidth is 200 GHz at 4.9 K.
	}

	\section{Introduction}
	Integrated electro-optic modulators are essential for high-bandwidth optical links in cryogenic environments, in applications such as delivering control signals and reading out data in solid-state quantum computing \cite{decea2020readout, youssefi2020cryogenic, veldhorst2017silicon, silverstone2016silicon, maurand2016cmos, holmes2013energy} or inter-satellite optical communications \cite{lange2009high, toyoshima2008ground}. Existing electro-optic modulators, however, suffer from low bandwidth or reduced modulation efficiency at cryogenic temperatures because they rely on tuning mechanisms that degrade with decreasing temperature. Devices based on free carriers such as silicon modulators \cite{gehl2017operation}, for example, suffer from carrier freeze-out at low temperature \cite{lengeler1974semiconductor}. This may be mitigated with degenerate doping but at the cost of increased insertion loss. Non-carrier based modulators using Franz-Keldysh or quantum confined Stark effect \cite{pintus2019characterization} or Pockels effect \cite{eltes2019integrated} suffer from weak electro-optic strength at low temperature, thus requiring higher drive voltage and increased footprint, which limits the integration density of photonic integrated circuits for cryogenic applications.

	Graphene modulators are a promising alternative for low-temperature applications. They rely upon tuning graphene’s absorption through electrostatic gating \cite{wang2008gate}, a mechanism which does not suffer from degradation at low temperatures \cite{li2008dirac, bolotin2008temperature}. The intrinsic electronic mobility of graphene increases upon cooling \cite{chen2008intrinsic, zhu2009carrier, fratini2008substrate}, such that the speed, determined by the RC charging time, should not intrinsically degrade at low temperature. However, no work has demonstrated low temperature operation of graphene modulators to date. Here we demonstrate a high-bandwidth graphene electro-optic modulator at 4.9 K. The mobility of the graphene in these devices increases with decreasing temperature, leading to reduced device resistance and better RC-limited bandwidth of graphene modulators. Therefore, graphene enables high-speed integrated electro-absorption modulator that naturally exhibits high bandwidth at cryogenic temperature.

	\section{Graphene-silicon nitride electro-absorption ring modulator}
\begin{figure}[!ht]
		\centering
		\noindent\makebox[\textwidth]{\includegraphics[width=174mm]{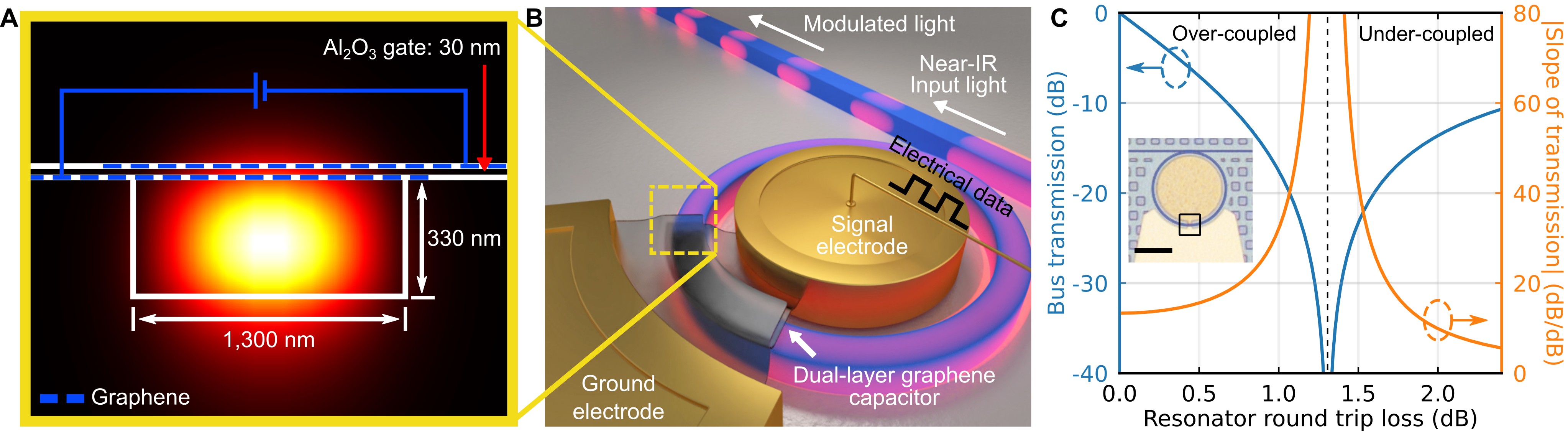}}
		\caption[]{\sf Graphene electro-absorption modulator design and resonator loss modulation at critical coupling.\\
			(A) The graphene modulator consists of a dual-layer graphene capacitor on top of a silicon nitride (\ch{Si3N4}) waveguide. The capacitor consists of two graphene sheets (blue dashed lines) separated by a 30 nm alumina (\ch{Al2O3}) gate dielectric (white solid lines are boundaries for \ch{Al2O3} gate dielectric and \ch{Si3N4} waveguide). The top graphene sheet is cladded with 40 nm \ch{Al2O3} (not shown in the cross-section). The \ch{Si3N4} waveguide (1,300 nm by 330 nm) is designed to ensure significant overlap between the fundamental quasi-TE mode and graphene capacitor via evanescent wave. By applying voltage between the graphene sheets, we electrostatically gate them and induce Pauli-blocking, suppressing interband transitions of carriers in graphene sheets and reducing optical absorption and mode propagation loss. (B) The graphene capacitor is embedded in a ring resonator to enhance graphene-light interaction while reducing footprint and capacitance for maximum bandwidth. (C) Simulated transmission (blue curve, left axis) and slope of transmission (orange curve, right axis) with respect to resonator round trip loss. To achieve strong modulation even with small voltage swing and small capacitance, we modulate resonator loss near critical coupling shown as dashed vertical line. The transmission is most sensitive (i.e. largest slope) when the resonator is near critical-coupling. We, therefore, design the resonator-bus coupling gap (around 180 nm) to be near critical-coupling. The modulator transitions from being critically-coupled to over-coupled as graphene absorption is modulated from high to low, respectively. Inset: An optical micrograph of the fabricated device showing the waveguide and ring resonator, false-colored in blue. The two electrodes for the graphene capacitor are shown in yellow. Boxed region indicates where the graphene capacitor is placed with 5 $\upmu$m device length around the ring with 40 $\upmu$m radius. Squares around the device are fill patterns for chemical mechanical planarization. Scale bar, 40 $\upmu$m.}
		\label{fig:schematics}
	\end{figure}

	The graphene electro-absorption modulator consists of a dual-layer graphene capacitor integrated with a silicon nitride (\ch{Si3N4}) waveguide. The capacitor consists of two graphene sheets (blue dashed lines in \autoref{fig:schematics}A) separated by 30 nm alumina (\ch{Al2O3}) gate dielectric (white solid lines in \autoref{fig:schematics}A are boundaries of \ch{Al2O3} gate dielectric and \ch{Si3N4} waveguide). The \ch{Si3N4} waveguide (1,300 nm by 330 nm) is designed to ensure significant overlap between the fundamental quasi-TE mode and graphene capacitor via evanescent wave as shown in \autoref{fig:schematics}A. By applying voltage to the graphene capacitor, we electrostatically gate the graphene sheets and induce Pauli-blocking, i.e. reduce optical absorption and mode propagation loss by suppressing interband transitions of carriers in graphene \cite{liu2011graphene}.

	We embed the graphene capacitor/\ch{Si3N4} waveguide in a ring resonator as shown in \autoref{fig:schematics}B to enhance graphene-light interaction while reducing footprint and capacitance for maximum bandwidth. In addition, we utilize resonator loss modulation at critical coupling \cite{yariv992585, phare2015graphene} to achieve strong modulation even with small voltage swing and small capacitance (about 9 fF, see the optical micrograph inset of \autoref{fig:schematics}C for device scale). \autoref{fig:schematics}C shows the simulated transmission through the waveguide (blue curve, left axis) and slope of transmission (orange curve, right axis) with respect to resonator round trip loss. The transmission is most sensitive (i.e. largest slope) when the resonator is near critical-coupling (i.e. near dashed vertical line in \autoref{fig:schematics}C). We therefore design the resonator-bus coupling gap (around 180 nm) to be near critical-coupling to achieve highest sensitivity to changes in graphene absorption, and to transition from being critically coupled to over-coupled as graphene absorption is modulated from high to low, respectively.

	\section{Results}
\begin{figure}[!ht]
		\centering
		\includegraphics[width=110mm]{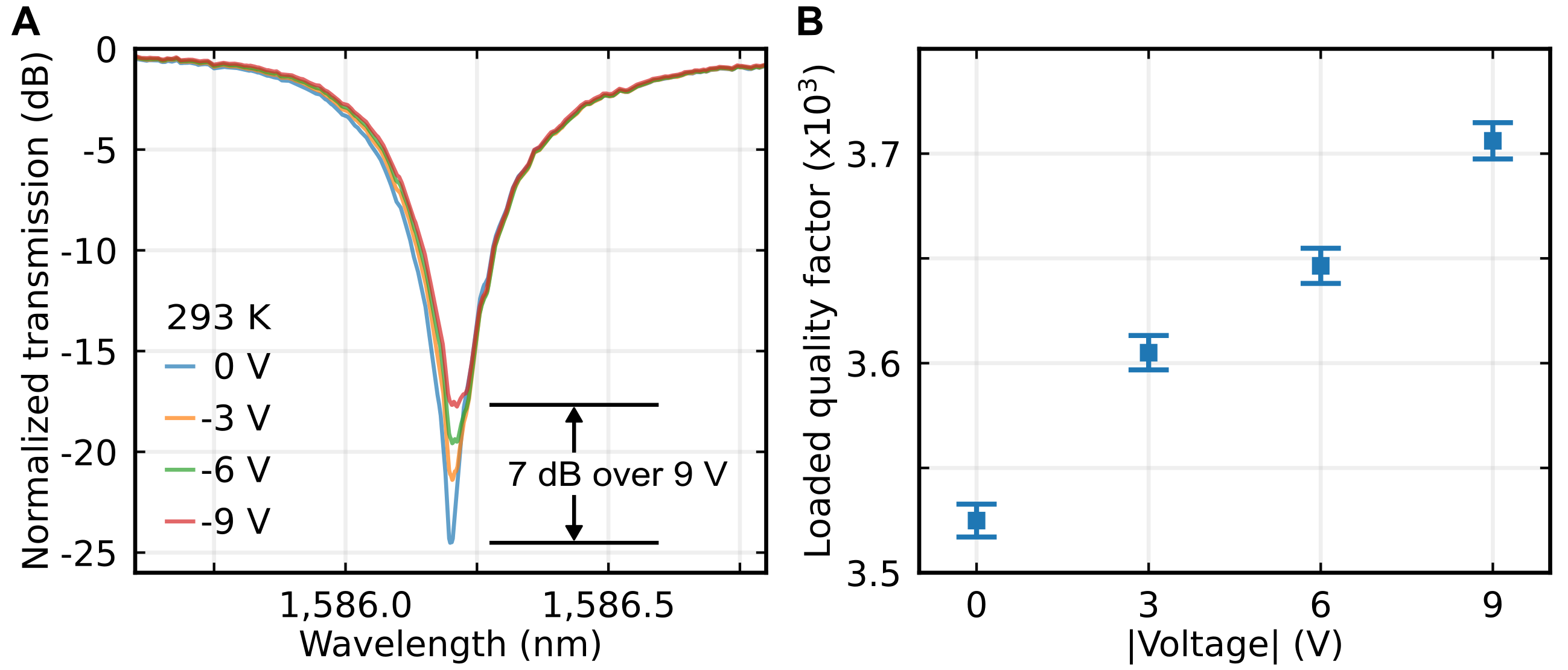}
		\caption[]{Modulating resonator-bus coupling condition by electrostatically gating the graphene capacitor.\\
			(A) Transmission spectra of the modulator at various d.c. voltages at room temperature. With applied voltage to the graphene modulator, we change the transmission at resonance by more than 7 dB by decreasing graphene absorption and changing the resonator-bus coupling condition. By reducing graphene absorption and resonator round trip loss, we make the resonator more over-coupled as indicated by the increasing transmission at resonance. (B) We measure the loaded quality factor $ Q_L $ of the resonator spectra in \autoref{fig:dc}A. The $ Q_L $ increases from about 3,500 to 3,700 over 9 V, corresponding to resonator round trip loss decreasing from 1.10 dB to 0.96 dB due to decreasing graphene absorption.}
		\label{fig:dc}
	\end{figure}

	We show modulation of transmission by more than 7 dB at room temperature by electrostatically gating the graphene and, thus, modulating the resonator round trip loss near critical coupling. In \autoref{fig:dc}A, we show the transmission spectra of the graphene ring modulator at different applied voltages. The transmission is about $-$25 dB at resonance (around 1,586.2 nm) when applied voltage is 0 V, indicating near critical-coupling of the ring with high graphene absorption. With applied voltage to the graphene modulator, the resonator-bus coupling condition becomes more over-coupled and the transmission at resonance increases. To measure the change in resonator round trip loss as a function of voltage, we measure the loaded quality factor $ Q_L $ from each of the spectra shown in \autoref{fig:dc}A and plot it in \autoref{fig:dc}B. The $ Q_L $ increases from about 3,500 to 3,700 over 9 V, corresponding to resonator round trip loss decreasing from 1.10 dB to 0.96 dB \cite{bogaerts2012silicon} due to decreasing graphene absorption.

\begin{figure}[!ht]
		\centering
		\includegraphics[width=82mm]{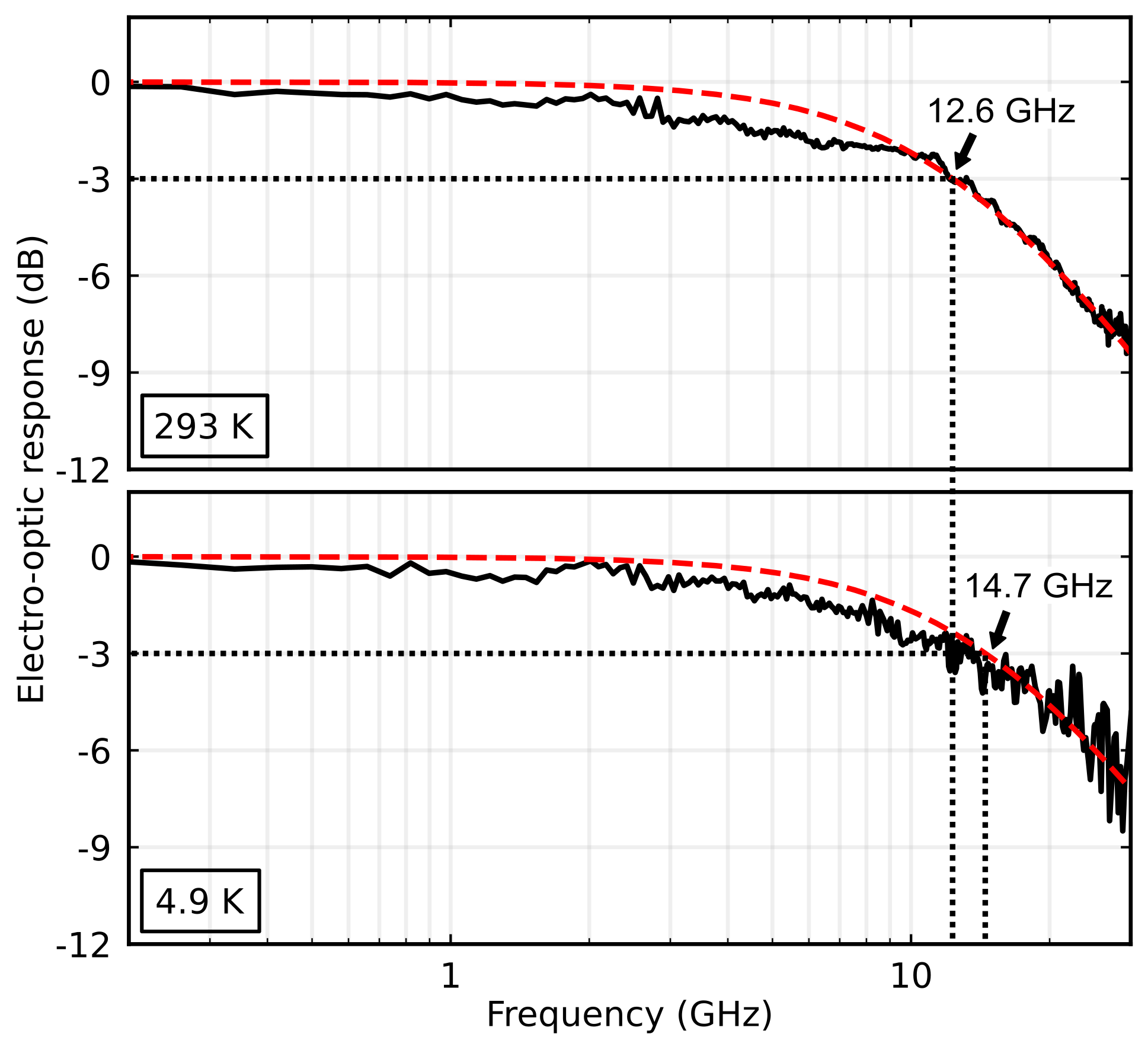}
		\caption[]{Enhancement of electro-optic bandwidth of graphene modulator at cryogenic temperature.\\
			The measured electro-optic bandwidth at 293 K and 4.9 K are 12.6 GHz and 14.7 GHz, respectively. Red dashed lines are single pole fitting to the data. The electro-optic response increases by about 16 \% when the device is cooled from room to cryogenic temperature. The modulator is driven with a vector network analyzer (see \autoref{sfig:setup} in the Supplementary Information) with RF power 13.5 dBm ($ \text{V}_{\text{pp}} $ = 3 V) at d.c. bias of -9 V at both temperatures.}
		\label{fig:ac}
	\end{figure}

	We characterize the frequency response of the graphene modulator at 293 K and 4.9 K and measure a 3-dB bandwidth of 12.6 GHz and 14.7 GHz, respectively (\autoref{fig:ac}). We place the graphene modulator in a cryogenic probe station to precisely control the temperature of the chip and perform high-speed electro-optic measurements (see \autoref{sfig:setup} for experimental setup at cryogenic temperature). The bandwidth increases approximately 16 \% when the device is cooled from room to cryogenic temperature despite being driven with the same d.c. bias (-9 V) and RF power (13.5 dBm, $ \text{V}_{\text{pp}} $ = 3 V, see \autoref{sfig:eye} for eye diagrams measured at 293 K in the Supplementary Information). We normalize the amplitude of each curve to the response at 1 MHz and fit it to a single pole transfer function $ 1/(1 + j2\pi f \tau) $, where $ f $ is the frequency and $ \tau $ is the modulator time constant. The 3-dB bandwidth at each temperature is measured from the fitted curve (shown as dashed lines in \autoref{fig:ac}) as $ f_{3dB}=1/(2\pi\tau) $.

	We show that the increase of graphene modulator's bandwidth at low temperature follows the trend of graphene carrier mobility. We pattern graphene test devices into Hall bar configuration, allowing extraction of the carrier mobility by simultaneous measurement of the four-terminal conductance and carrier density through the Hall effect (see \autoref{sfig:hallbar} in the Supplementary Information). As shown in \autoref{fig:mobility}A, the mobility increases from approximately 1,420 cm$^\text{2}$/Vs at 293 K to 1,650 cm$^\text{2}$/Vs at 4.9 K. This increase reflects the weakening of scattering from phonons in the graphene and the surrounding \ch{Al2O3} dielectric. The solid line in \autoref{fig:mobility}A shows a fit to the data which combines temperature-independent scattering from disorder (such as trapped charge in the \ch{Al2O3}) and temperature-dependent scattering from graphene longitudinal acoustic phonons and \ch{Al2O3}-graphene surface polar phonons \cite{chen2008intrinsic,zhu2009carrier,fratini2008substrate, fischetti2001effective} (see \autoref{sfig:mobility} in the Supplementary Information). In order to confirm that the device’s resistance governs the bandwidth change with temperature, we verify that the capacitance remains constant with temperature. The capacitance changes by less than 5 \% from 293 K to 1.5 K by measuring the change in carrier concentration of the graphene Hall bar at various temperatures (see \autoref{sfig:carrier_temp} in the Supplementary Information).

	\begin{figure}[!ht]
		\centering
		\includegraphics[width=122mm]{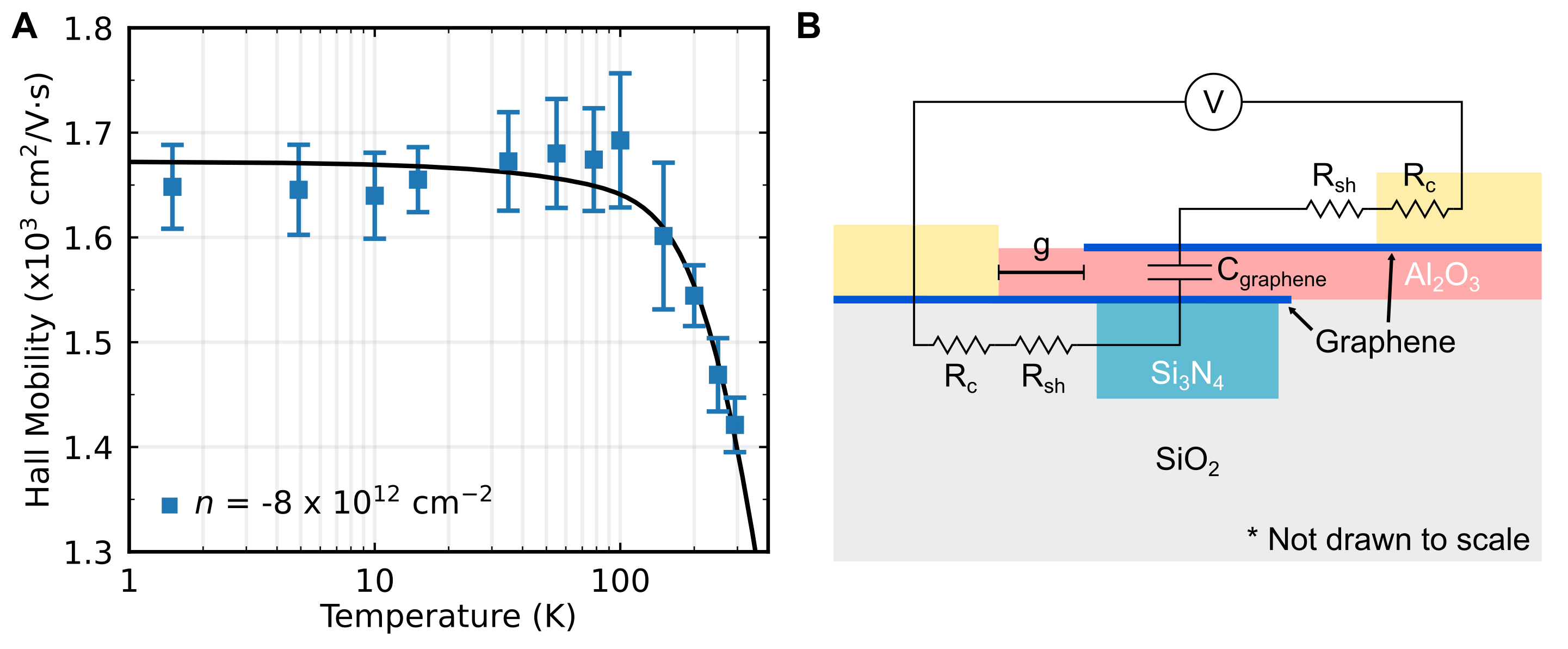}
		\caption[]{Mobility enhancement of graphene at cryogenic temperature and its correspondence to graphene modulator.\\
			(A) We extract graphene mobility using a Hall bar (see \autoref{sfig:hallbar} in the Supplementary Information) with respect to temperature. We measure an increase of mobility from approximately 1,420 cm$^\text{2}$/Vs at 293 K to 1,650 cm$^\text{2}$/Vs at 4.9 K. This increase reflects the weakening of scattering from phonons in the graphene and the surrounding \ch{Al2O3} dielectric. The black solid line shows a fit to the data which combines temperature-independent scattering from disorder (such as trapped charge in the \ch{Al2O3}) and temperature-dependent scattering from graphene longitudinal acoustic phonons and \ch{Al2O3}-graphene surface polar phonons (see \autoref{sfig:mobility} in the Supplementary Information). The measured mobility follows the temperature dependence of the fitted curve well within the measurement error. (B) The equivalent RC circuit of the graphene modulator is overlaid with the device cross-section. $ R_c $ is the graphene contact resistance, $ R_{sh} $ is the graphene sheet resistance, $ C $ is the dual-layer graphene capacitor, and $ g $ is the gap between the electrode and the capacitor. The enhancement of graphene carrier mobility at cryogenic temperature reduces the sheet resistance of the device, increasing the RC-limited bandwidth. 
		}
		\label{fig:mobility}
	\end{figure}

	\section{Discussion}
	Using the measured mobility and two-terminal device resistance, we determine that the graphene modulator bandwidth is currently limited by its high contact resistance, whereas its high mobility (and resulting low sheet resistance) at cryogenic temperature supports a fundamental bandwidth of 200 GHz. From the RC circuit in \autoref{fig:mobility}B, the modulator's total RC-limited bandwidth is,
	\begin{equation}
	f_{3dB} = \frac{1}{2\pi(2R_c + 2R_{sh}g)C},
	\label{eq:bw}
	\end{equation}
	where $ R_c $ is the contact resistance, $ R_{sh} $ is the sheet resistance, $ g $ is the gap between the electrodes and capacitor, and $ C $ is the capacitance per length. The total bandwidth is governed by contributions from extrinsic (i.e. due to parasitic components) and intrinsic bandwidth:
	\begin{equation}
	\begin{split}
	1/f_{3dB} &= 1/BW_{ext} + 1/BW_{int}\\
	&= 2\pi (2R_c C) + 2\pi(2R_{sh}gC),
	\label{eq:bw2}
	\end{split}
	\end{equation}
	where $ BW_{ext} = 1 / (2\pi [2R_cC]) $ is the extrinsic bandwidth governed by parasitic contact resistance and $ BW_{int} = 1 / (2\pi [2R_{sh}gC]) $ is the intrinsic device bandwidth governed by the capacitance and sheet resistance. From Hall bar measurements (see \autoref{sfig:hallbar} in the Supplementary Information) and using \autoref{eq:bw}, we extract $ C $ and $ R_{sh} $ at 4.9 K as 1.85 fF/$ \upmu $m and 470 $ \Upomega $/sq, respectively, which translates to $ BW_{int} = \text{200 GHz}$. With state-of-the-art graphene contact resistance measured as low as 100 $ \Upomega \upmu$m \cite{leong2014low}, we expect to enhance the extrinsic bandwidth and achieve a total bandwidth of $ > $ 130 GHz at cryogenic temperature with optimized graphene contacts without changing the device cross-section. Note that this high bandwidth requires ensuring that the cavity photon lifetime is not a limiting factor, which can be done by, for example, increasing the graphene capacitor length around the circumference of the ring resonator from 5 $ \upmu $m to 33 $ \upmu $m (see Supplementary Information).

	\section{Conclusion}
	We have demonstrated high-speed integrated graphene modulator with measured bandwidth of 14.7 GHz with intrinsic bandwidth of 200 GHz at 4.9 K by leveraging graphene carrier mobility improvement at low temperature. In contrast to traditional electro-optic tuning mechanisms that decrease in bandwidth or modulation efficiency at low temperature, the graphene modulator exhibits an increase in electro-optic response from 293 K to 4.9 K without trading off voltage or footprint. By optimizing graphene contact resistance and device footprint, we expect the graphene modulator to support bandwidths $ > $ 130 GHz at cryogenic temperature. In addition, with optimized quality of gate dielectric, the modulation extinction ratio, currently limited by weaker dielectric constant and breakdown field of the \ch{Al2O3} gate compared to values reported in literature \cite{yota2013characterization}, could be further increased (see \autoref{sfig:design} in the Supplementary Information). This natural enhancement of electro-optic response at low temperature makes graphene modulators versatile and suitable for high-speed electro-optic applications at cryogenic temperature.

	\section*{Methods}
	Please see the Supplementary Information for detailed description of device fabrication and experimental measurements.

	\section*{Acknowledgement}
	The authors would like to thank Dr. Min Sup Choi, Dr. Christopher T. Phare, Dr. Andres Gil-Molina, Dr. Nathan C. Abrams, Ipshita Datta, Min Chul Shin, and Euijae Shim for the fruitful discussions and experimental support. This work was performed in part at the City University of New York Advanced Science Research Center NanoFabrication Facility and in part at the Columbia Nano Initiative (CNI) shared labs at Columbia University in the City of New York.

	\section*{Funding}
	We also gratefully acknowledge support from the Office of Naval Research for award \#N00014-16-1-2219, Defense Advanced Research Projects Agency program for award \#HR001110720034, National Science Foundation for award \#UTA16-000936, National Aeronautics and Space Administration for award \#NNX16AD16G, Air Force Office of Scientific Research for award \#FA9550-18-1-0379, Air Force Materiel Command for award \#FA8650-18-1-7815, and Hypres, Inc. for award \#CU15-3759.

	\bibliographystyle{unsrt}
	\bibliography{References}

	\clearpage
	\setcounter{page}{1}
	\setcounter{figure}{0}
	\setcounter{section}{0}
	\setcounter{equation}{0}
	
	\renewcommand{\thepage}{S\arabic{page}}
	\renewcommand{\thefigure}{S\arabic{figure}}
	\def\theequation{S\arabic{equation}}

	
	\begin{center}
		\Huge \textbf{Supplementary Information}
	\end{center}
	
	\section{Bandwidth with optimized graphene contact resistance and adjusted cavity quality factor}
	In the circuit diagram in \autoref{fig:mobility}B, we assume negligible pad and graphene quantum capacitance as the graphene capacitance is in the order of 9 fF, dominating other series capacitances. The RC-limited bandwidth is then,
	\begin{equation}
	f_{3dB} = \frac{1}{2\pi RC} = \frac{1}{2\pi \left( 2\frac{R_c}{l} + 2R_{sh}\frac{g}{l} \right) \epsilon_0 k \frac{wl}{d}},
	\label{seq:rc_bandwidth}
	\end{equation}
	where $ R_c $ is the graphene contact resistance (\si{\ohm\micro\meter}), $ R_{sh} $ is the sheet resistance (\si{\ohm}/sq), $ g $ is the gap between graphene capacitor and electrode (see \autoref{fig:mobility}), $ l $ is the device length around the ring resonator circumference, $ \epsilon_0 $ is the permittivity of free-space, $ k $ is the dielectric constant, $ w $ is the capacitor width (i.e. graphene overlap width), and $ d $ is the gate dielectric thickness. The length contribution in both sheet and contact resistances cancels out the capacitance length, so \autoref{seq:rc_bandwidth} simplifies to,
	\begin{equation}
	f_{3dB} = \frac{1}{2\pi(2R_C + 2R_{sh}g) \epsilon_0 k \frac{w}{d}}.
	\label{seq:rc_bandwidth_reduced}
	\end{equation}
	Therefore, the device length does not impact the RC bandwidth but only the resonator round trip loss. If we optimize graphene contact resistance close to the state-of-the-art value of 100 \si{\ohm\micro\meter} \citeSup{supleong2014low} while keeping other values constant (i.e. sheet resistance, capacitance, gate dielectric constant, etc, same as the current device), the 3-dB bandwidth is approximately 137 GHz at 4.9 K. This bandwidth exceeds the current photon lifetime of the resonator (with Q $\approx$ 3,700 in \autoref{fig:dc}B, the resonator linewidth is $\approx$ 52 GHz). We can mitigate this by adjusting the resonator round trip loss by increasing the device length as it does not affect the RC bandwidth as shown in \autoref{seq:rc_bandwidth_reduced}. To ensure that the cavity photon lifetime supports $ > $ 130 GHz, the loaded quality factor needs to be around 1,490, corresponding to resonator round trip loss of 6.4 dB, or 33 \si{\micro}m device length with current graphene absorption.

	\section{Methods}
	\subsection{\ch{Si3N4} ring resonator fabrication}
	We deposited a film of 330 nm-thick low-pressure chemical vapor deposition (LPCVD) \ch{Si3N4} on top of 4.3 \si{\micro\meter}-thick thermally grown \ch{SiO2} on silicon substrate. We spin-coated Ma-N 2403 resist and exposed ring resonators with 1.3 \si{\micro}m width and 40 \si{\micro}m radius using e-beam lithography (Elionix ELS-G100, 100 kV). We ensured only the fundamental quasi-TE mode propagated in our devices by designing an inversely tapered waveguide at the input and output facets. Before developing the resist, we exposed the chemical mechanical planarization (CMP) fills (square patterns around the device in the inset of \autoref{fig:schematics}C) using a 248 nm wavelength Deep-UV (DUV) mask aligner (MA6-SUSS) so that the waveguides and fills were patterned on the same \ch{Si3N4} layer. We took advantage of Ma-N resist's sensitivity to DUV wavelengths to expose CMP fills because direct writing with e-beam lithography would have taken considerable amount of writing time. After exposing the waveguides and CMP fills, the resist was developed and etched using inductively-coupled plasma etcher with \ch{CH3}/\ch{O2} etching gas. The patterned \ch{Si3N4} layer was cladded with \ch{SiO2} by plasma-enhanced chemical vapor deposition (PECVD). The cladded surface was planarized using CMP so that the transferred graphene sheets would not rupture from the steps in the top cladding caused by the waveguide patterns. The planarized surface was cleaned with RCA-2 (\ch{H2O2}/\ch{NH4OH}/\ch{H2O}) to remove slurry residues from the CMP.

	\subsection{Graphene transfer through electrochemical delamination}
	We used CVD graphene grown by Grolltex Inc. \citeSup{supgrolltex}. We spin-coated PMMA 495K A6 at 2,000 RPM on graphene grown on Cu substrate. This gave a 400 nm-thick polymer support for the graphene transfer. We dried the spin-coated PMMA on graphene overnight at ambient conditions without additional baking. We electrochemically delaminated the PMMA/graphene stack from the Cu substrate as described in Ref. \citeSup{supyibo.nanolett.8b01091}. We prepared 1M \ch{NaOH} solution as the electrolyte with a PMMA/graphene/Cu foil as the cathode, and a bare Cu foil as the anode. With applied voltage of around -2.2 V between the two terminals, a 20 mm x 20 mm PMMA/graphene film delaminated from the Cu substrate and floated on the electrolyte surface within few tens of seconds. The delaminated PMMA/graphene stack was transferred to a fresh DI water bath with a glass slide and transferred onto \ch{O2} plasma-treated substrate (i.e. on top of planarized waveguide surface). The \ch{O2} plasma removed any residues on the substrate and made the surface hydrophilic, which facilitated the removal of trapped water residue in the drying step. The substrate with transferred PMMA/graphene stack was put into a vacuum desiccator with a base pressure of around 0.5 Torr for at least 24 hours to pump out residual water trapped between the stack and substrate during the wet transfer. Fully removing the water residue was critical to reducing tears and damages to the transferred graphene as residual water can rupture graphene during the baking step. After vacuum-drying residual water, the substrate was baked on a hot plate for 2 hours at 180 \si{\celsius}. Lastly, PMMA was dissolved using acetone for at least 1 hour.

	\subsection{Dual-layer graphene capacitor fabrication}
	To pattern the graphene layer after transfer, we used a two-layer resist mask using PMMA 495K A2 at 1000 RPM (around 100 nm) and hydrogen silsesquioxane (HSQ, XR-1561 6 \%) at 2,000 RPM (around 130 nm) on the top. We noticed polymer based negative resist, such as Ma-N, leaves residues that are difficult to selectively clean with respect to graphene. Using an inorganic resist, such as HSQ, with a sacrificial layer left minimal residues on graphene. We patterned the graphene sheet 5 \si{\micro}m long around the arc of the \ch{Si3N4} ring resonators with e-beam lithography. We etched through the PMMA mask and graphene with reactive ion etching (RIE) \ch{O2} plasma. Once the graphene was etched, we stripped the resist in acetone, where it dissolved the bottom PMMA and cleanly removed the HSQ mask. After patterning the graphene sheet, we defined the metal electrode 0.45 \si{\micro}m away from the waveguides with bilayer PMMA resist (100 nm-thick PMMA 950k A2 on top of 500 nm-thick PMMA 495K A6) and e-beam lithography, followed by the deposition and lift-off of 1 nm/45 nm/55 nm thick Cr/Pd/Au layers at high vacuum ($< \text{10}^{-\text{7}}$ Torr). We placed the electrodes close to the waveguide in order to maximize bandwidth. Eigenmode simulations showed mode loss due to electrodes placed in close proximity to the waveguide (21 dB/cm) was still an order of magnitude lower than that from graphene absorption even at its maximum expected transparency (683 dB/cm). After metal deposition, 1 nm of Al was thermally evaporated to provide nucleation seed layer for the following ALD deposition of \ch{Al2O3} gate dielectric (30 nm thickness). Second layer graphene was transferred on top of this \ch{Al2O3} gate dielectric as outlined in the transfer section. Graphene patterning and electrode deposition steps were repeated for the second graphene sheet. After fabricating the Gr/\ch{Al2O3}/Gr capacitor, a final layer of 40 nm-thick \ch{Al2O3} was deposited to clad and protect the top graphene layer. Vias were etched with buffered oxide etch (50:1 ratio) to remove \ch{Al2O3} on top of the electrodes to access the metal contacts.

	\subsection{High-speed electro-optic measurements at cryogenic temperature}
\begin{figure}[!htp]
		\centering
		\noindent\makebox[\textwidth]{\includegraphics[width=183mm]{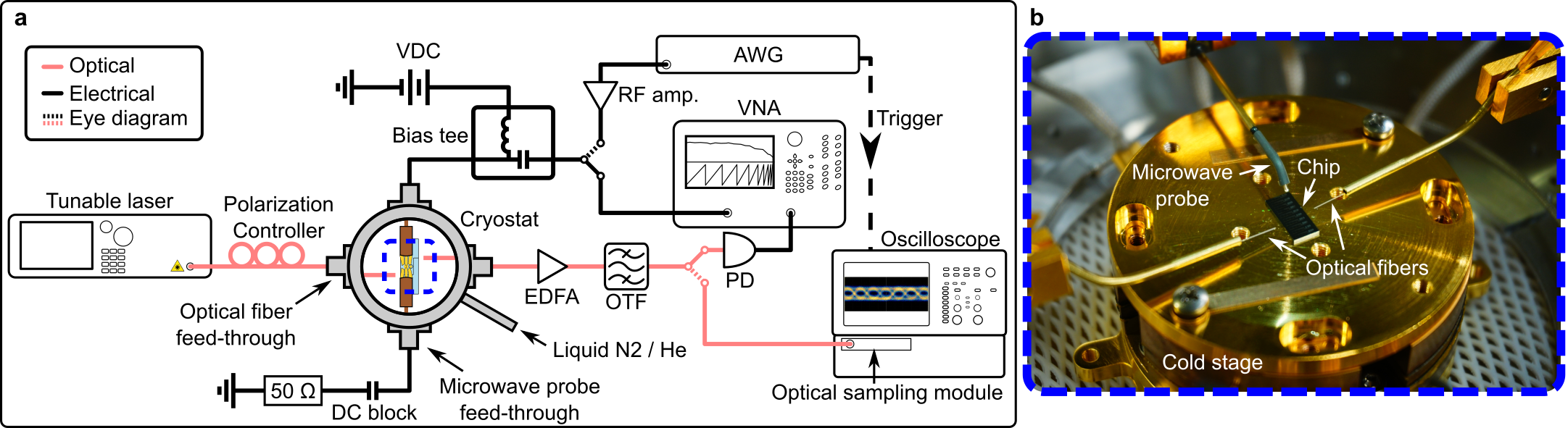}}
		\caption[]{Experimental setup for high-speed electro-optic measurements.\\
			(a) Schematics of high-speed electro-optic experiment at cryogenic temperature. The graphene modulator was placed inside the cryogenic probe station (Janis Research ST-500) equipped with optical fiber and microwave probe feed-throughs. We couple input light from a tunable laser into the chip by edge-coupling the optical fiber  to the inverse tapered waveguide. We control the polarizaion of the input light using a polarization controller so that it is closely matched to the waveguide fundamental quasi-TE mode. The output light is collected with another output fiber at the opposite end. It is then amplified through an erbium-doped fiber amplifier (EDFA) to compensate for fiber-to-chip coupling loss. The light is then passed through an optical tunable filter (OTF) with a 0.25 nm bandwidth to increase signal-to-noise ratio by suppressing amplified spontaneous emission (ASE) noise from the EDFA. To measure the electro-optic response with respect to frequency, we excited the modulator with electrical signal from the vector network analyzer (VNA). The VNA sweeps the frequency with input electrical power of 13.5 dBm ($V_{pp} = 3$~V) and fed into a bias tee, where a constant d.c. bias of -9 V is added to the RF signal. The IF filter bandwidth of the VNA was kept at 1 kHz for fast sweeps. The electrical signal path length was much shorter than the IF filter dwell time for accurate sweeps. The combined a.c. and d.c. electrical signal is applied to the modulator through a GSG microwave (67 GHz bandwidth). Another GSG microwave probe is landed on the modulator where it terminates the modulator with a 50~\si{\ohm} termination and a d.c. block to minimize electrical reflection due to large impedance mismatch between the modulator and transmission line. The output optical signal is received by a high-speed photodetector and input into the second port of the VNA. To acquire eye diagrams shown in \autoref{sfig:eye}, VNA was replaced with an arbitrary waveform generator (AWG) that output non-return-to-zero (NRZ) $ 2^9 -1 $ pseudo-random binary stream. The electrical output from the AWG was amplified to $ V_{pp} = \SI{3}{\volt}$ by an RF amplifier and fed into the bias tee to add the d.c. bias. For the output, the output optical fiber after the OTF was fed directly into the optical sampling module in the digital communication analyzer oscilloscope to acquire the eye diagram. (b) Photograph of the cold stage inside the cryogenic probe station. The chip was affixed with the holders on the cold stage, which was cooled with the liquid cyrogens. The two optical fibers coupled light into and out of the chip. A microwave GSG probe drove electrical signal to the graphene modulator. The second GSG prove used to terminate the modulator is not shown.}
		\label{sfig:setup}
	\end{figure}

\begin{figure}[!htp]
		\noindent\makebox[\textwidth]{\includegraphics[width=183mm]{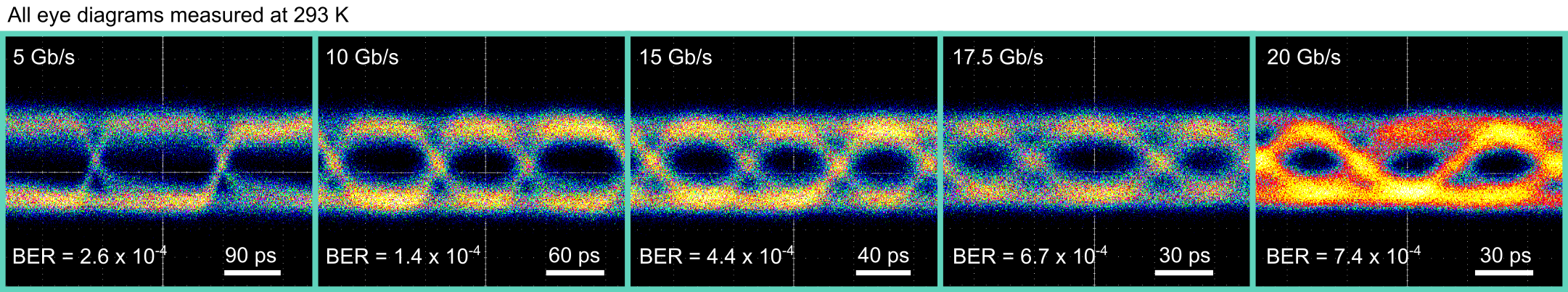}}
		\caption[]{Eye diagrams of non-return-to-zero (NRZ) $ 2^9 - 1 $ pseudo-random binary stream at 293 K. \\
			The eye diagrams of graphene modulator with data transmission speeds from 5 Gb/s to 20 Gb/s were acquired at 293 K with a drive voltage $V_{pp} = 3$~V at d.c. bias of -9 V. The bit-error rate (BER) for each data rate was calculated from the measured quality factor (Q) of the eye diagram using the relation BER$ = 0.5\text{~erfc}(Q/\sqrt{2}) $, where erfc is the complementary error function \citeSup{supagrawal2012fiber}.}
		\label{sfig:eye}
	\end{figure}

	Electro-optic measurements were conducted in a cryogenic probe station (Janis Research ST-500) with two optical fiber feed-throughs for optical input and output as well as two microwave 67 GHz microwave GSG probes. \autoref{sfig:setup} describes the experimental setup. The sample stage was cooled down to 4.9 K using liquid helium. The stage temperature was closely monitored with a thermocoupler throughout the experiment to see any changes in temperature due to optical coupling and microwave excitation. Laser power was kept at 10 dBm maximum, where no signs of stage temperature fluctuation was observed. Fiber-to-chip coupling loss was around 10-12 dB per facet due to large modal mismatch between the cleaved single mode fiber inside the cryostat and on-chip inversely tapered waveguide couplers. For frequency response measurements, a microwave drive signal of 13.5 dBm ($\approx V_{pp} = 3 $~V) from the vector network analyzer (VNA, Anritsu MS4640B) was applied to a bias-tee that combined a d.c. bias of -9 V and applied to the modulator through a GSG probe. Another GSG microwave probe was landed on the modulator where it terminated the modulator with a 50~\si{\ohm} termination and a d.c. block to minimize electrical reflection due to large impedance mismatch between the modulator and transmission line. The output light was collected using an output fiber feed-through in the cryostat and fed into to an erbium-doped fiber amplifier (EDFA) to compensate for coupling and insertion losses followed by an optical tunable filter (OTF) with 1 nm bandwidth to reduce amplified stimulated emission (ASE) noise from the EDFA and increase optical signal-to-noise ratio. The filtered optical signal was fed into a high-speed photodetector (67 GHz Anritsu MN4765B), whose electrical signal was connected to the second port of the VNA. The IF bandwidth of the VNA was kept at 1 kHz for all measurements to ensure fast frequency sweeps but with dwell time much longer than the electrical path length for accurate measurements. The electrical paths including the GSG probes were de-embedded prior to the electro-optic experiment using a calibration substrate at relevant temperatures. To acquire eye diagrams for data communication at 293 K shown in \autoref{sfig:eye}, VNA was replaced with an arbitrary waveform generator (AWG, Keysight M8195A) and OTF filter was reduced to 0.25 nm bandwidth. AWG output non-return-to-zero (NRZ) $ \text{2}^\text{9}-\text{1} $ pseudo-random binary stream at various data rates from 5 Gb/s to 20 Gb/s. The output was fed into a RF amplifier to increase the drive voltage to $ V_{pp} = 3 $~V. The optical output signal was directly put into the oscilloscope's optical sampling module (Tektronix 80C02) instead of the high-speed photodetector to obtain the eye diagrams. The quality factor (Q) of the eye diagrams were measured, from which the bit-error rate (BER) was calculated from the relation BER$ = 0.5\text{~erfc}(Q/\sqrt{2}) $, where erfc is the complementary error function \citeSup{supagrawal2012fiber}.

	\subsection{Graphene Hall bar measurements for extracting mobility}
\begin{figure}[!htp]
		\centering
		\includegraphics[width=144mm]{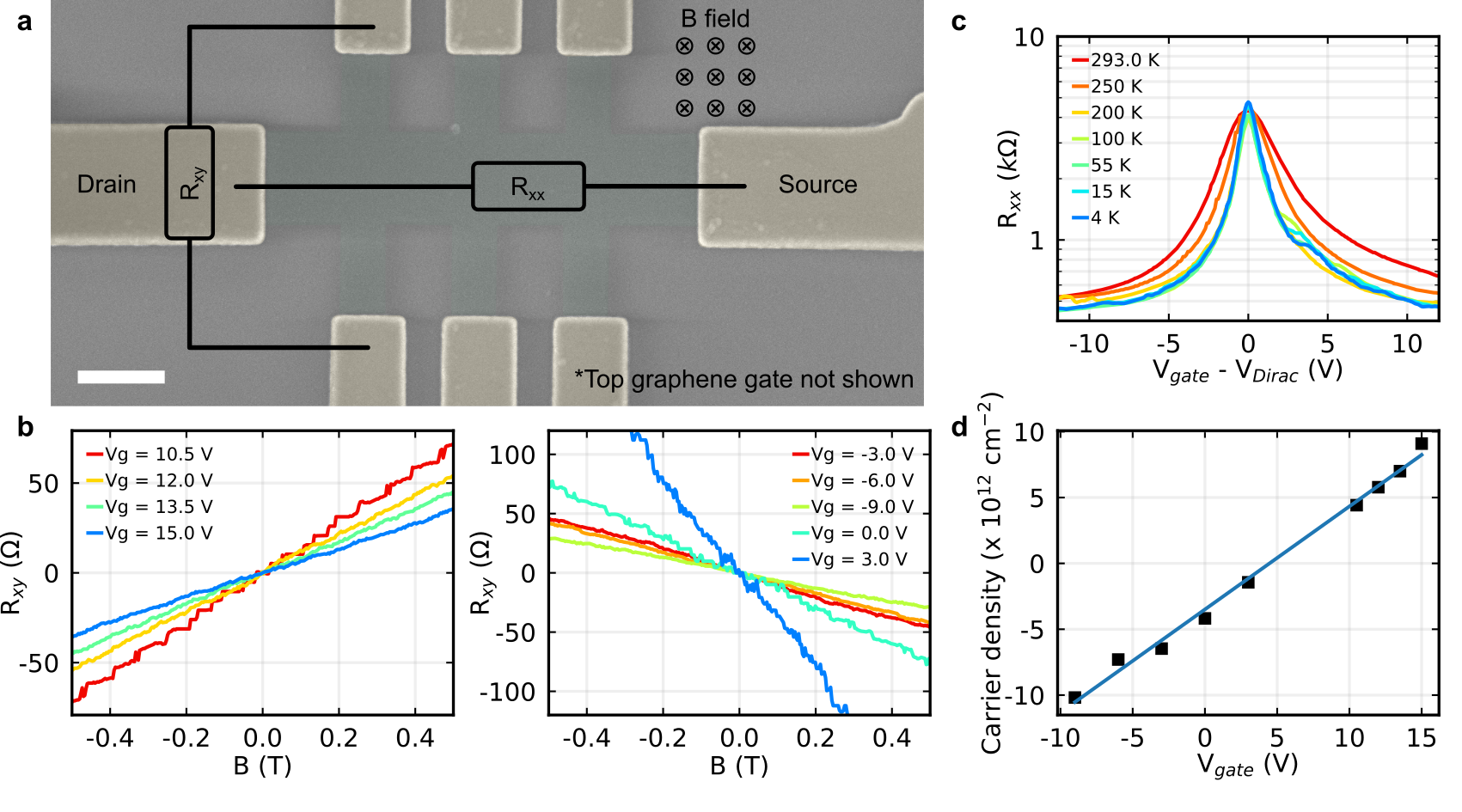}
		\caption[]{Graphene Hall bar transport measurements.\\
			(a) False-colored scanning electron microscope image of the graphene Hall bar used to measure CVD graphene mobility at various temperatures. The green shaded region is the bottom graphene sheet whose carrier density was controlled with a top graphene gate (not shown) overlapping the whole device, including the contacts. The yellow shaded region is the Cr/Pd/Au electrodes. Hall resistance ($ R_{xy} $) was measured by applying a current through drain-source (whose resistance is $ R_{xx} $) and measuring the potential difference between a pair of electrodes perpendicular to the drain-source while sweeping the magnetic field ($ B $) applied normal to the device plane. Scale bar, \SI{1}{\micro\meter}. (b) Hall resistance with respect to sweeping magnetic field at various gate voltages ($ V_g $). The left and right plot show gate voltages that electron and hole dope the graphene, respectively. The slopes of linear fits are used to extract carrier density at each gate voltage from the relation $ R_{xy} = B/ne $, where $ n $ is the carrier density and $ e $ is the elementary positive charge. The results are plotted in (d). (c) Sheet resistance $ R_{xx} $ of the drain-source channel with respect to gate voltage at various temperatures. These curves were used to extract Hall mobility of carrier concentration $ n = \SI{-8e12}{\per\centi\meter\squared} $ (corresponding to $ V_g = -9\si{\volt} $) at different temperatures from the relation $ R_{xx} = 1/\mu |n| e $. The extracted mobility with respect to temperature is plotted in \autoref{fig:mobility}A. (d) Carrier density with respect to gate voltage extracted from linear fits in (b). The blue line is the linear fit, which yields a capacitance of $ C = \SI{1.25}{\femto\farad\per\micro\meter\squared}$ from the relation $ n = CV/e $. The graphene modulator is biased at -9\si{\volt}, which corresponds to $ n=\SI{-8e12}{\per\centi\meter\squared} $. The dielectric constant of \ch{Al2O3} extracted from the fit was $ k= 4.2 $.}
		\label{sfig:hallbar}
	\end{figure}

\begin{figure}[!htp]
		\centering
		\includegraphics[width=144mm]{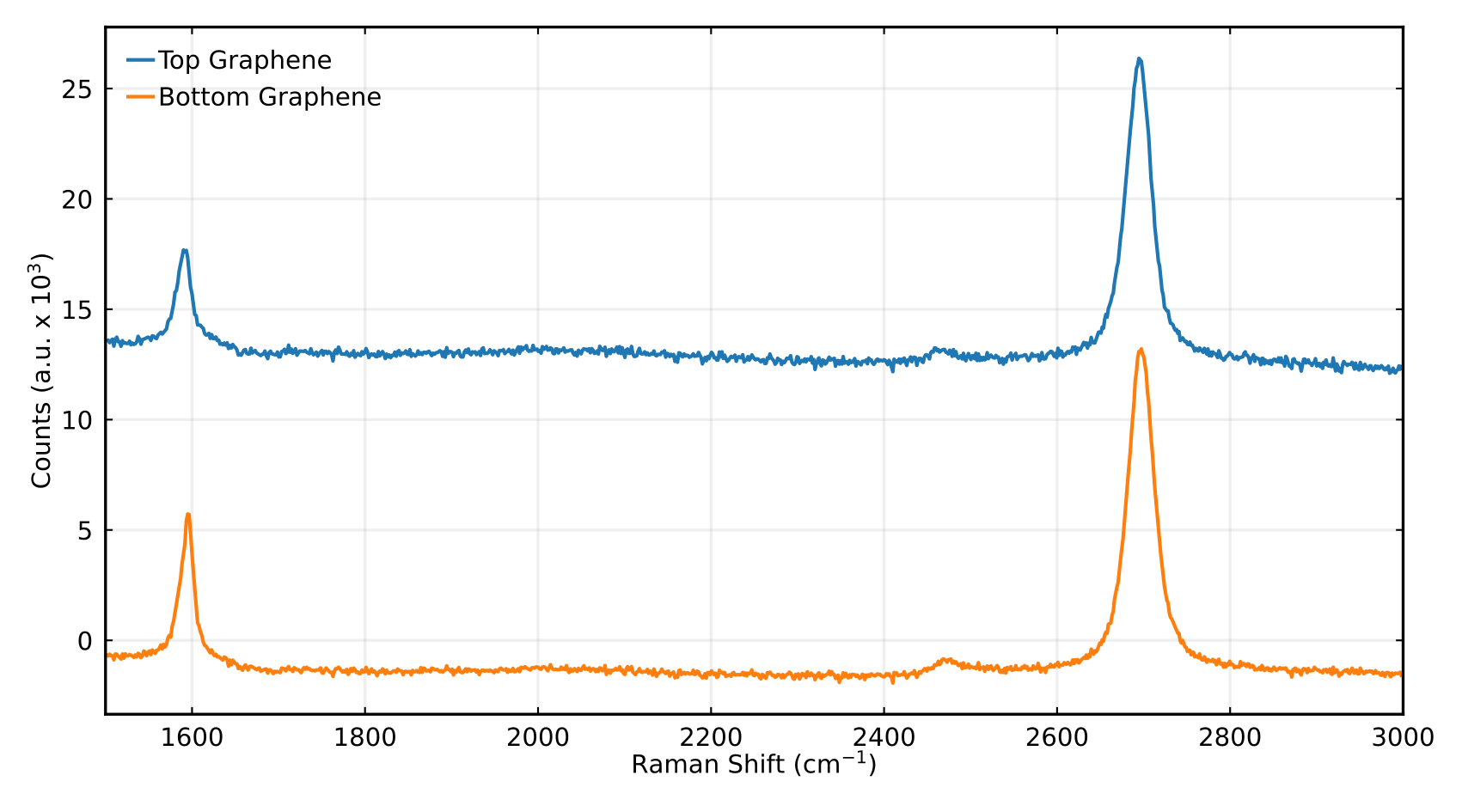}
		\caption[]{Raman spectroscopy of the two graphene sheets of the modulator.\\
			Graphene 2D peaks are at 2,697.4 \si{\per\centi\meter} and 2,694.6 \si{\per\centi\meter}, G peaks are at 1,595.0 \si{\per\centi\meter} and 1591.7 \si{\per\centi\meter}, and I(2D)/I(G) ratios are 1.71 and 1.93 for bottom and top graphene, respectively. The Raman spectroscopy measurements suggest an intrinsic carrier density of around \SI{-5e12}{\per\centi\meter\squared} \citeSup{supdas2008raman}, similar to that obtained from the Hall bar measurements (\SI{-4.1e12}{\per\centi\meter\squared}) in \autoref{sfig:hallbar}.}
		\label{sfig:raman}
	\end{figure}

	A Hall bar with similar cross-section (Gr/30 nm \ch{Al2O3}/Gr) as the modulator was fabricated on a similar substrate (330 nm LPCVD \ch{Si3N4}/4.3 \si{\micro}m \ch{SiO2}/Si) to extract the Hall mobility of graphene with respect to temperature. The bottom graphene was gated with a top graphene sheet overlapping the channel layer as well as the contact regions to ensure complete gating of the bottom channel. We put the graphene Hall bar in a closed-cycle refrigerator cryostat with a superconducting magnet. We measured the Hall resistance ($R_{xy}$) across a pair of contacts perpendicular to the drain-source channel (\autoref{sfig:hallbar}a) while sweeping the magnetic field from -0.5 to 0.5 T at various gate voltages, as plotted in \autoref{sfig:hallbar}b. The linear fit at each gate voltage provides carrier density versus gate voltage through the relation $ R_{xy} = B/ne $, where $R_{xy}$ is the measured Hall resistance, $B$ is the applied magnetic field, $n$ is the carrier density, and $e$ is the elementary positive charge. Doping carrier density is plotted against gate voltage in \autoref{sfig:hallbar}d, where intrinsic carrier density is measured to be around \SI{4.1e12}{\per\centi\meter\squared} corresponding to Dirac voltage of about 5 V, consistent with intrinsic carrier density deduced from Raman spectroscopy measurements (\autoref{sfig:raman}). We measured our dielectric constant to be $ k = 4.2 $ with breakdown field strength of about 5.6 MV/cm. The reduced gate dielectric quality could be due to defects from the graphene transfers, adversely affecting the dielectric characteristics \citeSup{supverweij1996dielectric}. We then measured sheet resistance of the graphene channel ($ R_{xx} $) while sweeping the gate voltage at various temperatures as shown in \autoref{sfig:hallbar}c. The Hall mobility was extracted from this plot using the relation $ R_{xx} =  1/\mu |n| e$, where $\mu$ is the mobility at doping concentration $n$. We used $n =- \SI{8e12}{\per\centi\meter\squared}$, corresponding to about $V_g = -9$~V, which is the d.c. bias applied to our graphene modulators in electro-optic measurements.

	\clearpage
	\section{Supplementary Figures}
	\vspace{10mm}

	\begin{figure}[!htp]
		\centering
		\includegraphics[width=70mm]{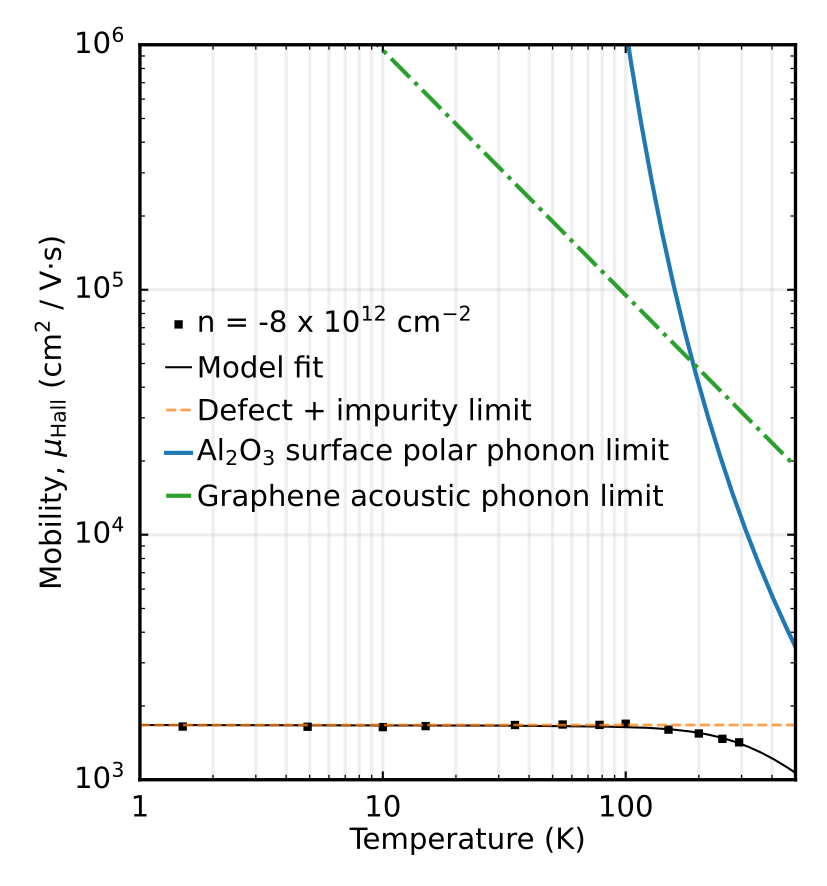}	\caption[]{Graphene mobility modeling.\\
			The measured mobility is fitted with the model described in Ref.\citeSup{supzhu2009carrier}. Square marks are experimental data from \autoref{fig:ac}B at carrier density \SI{-8e12}{\per\centi\meter\squared}. Dashed orange line is the defect and impurity limited mobility, which is temperature insensitive. Blue line is the mobility limited by surface polar phonon scattering from the \ch{Al2O3} gate dielectric. The material parameters for \ch{Al2O3} were referenced from Ref.\citeSup{supfischetti2001effective}. Green dotted dashed line is mobility limited by graphene acoustic phonon, described in Ref.\citeSup{supchen2008intrinsic}. Black solid line is the fit to data summing all the mobility limit contributions, from the relation $\mu_{total}^{-1} = \mu_C^{-1} + \mu_{SR}^{-1} + \mu_{GA}^{-1} + \mu_{SPP}^{-1}$, where $\mu_{total}$ is the total mobility, $\mu_{C}$ is the mobility limited by Coulomb scattering from impurities, $\mu_{SR}$ is the mobility limited by short-range scattering from defects, $\mu_{GA}$ is the mobility limited by graphene acoustic phonon, and $\mu_{SPP}$ is the mobility limited by substrate surface polar phonon scattering.}
		\label{sfig:mobility}
	\end{figure}

	\begin{figure}[!htp]
		\centering
		\includegraphics[width=61mm]{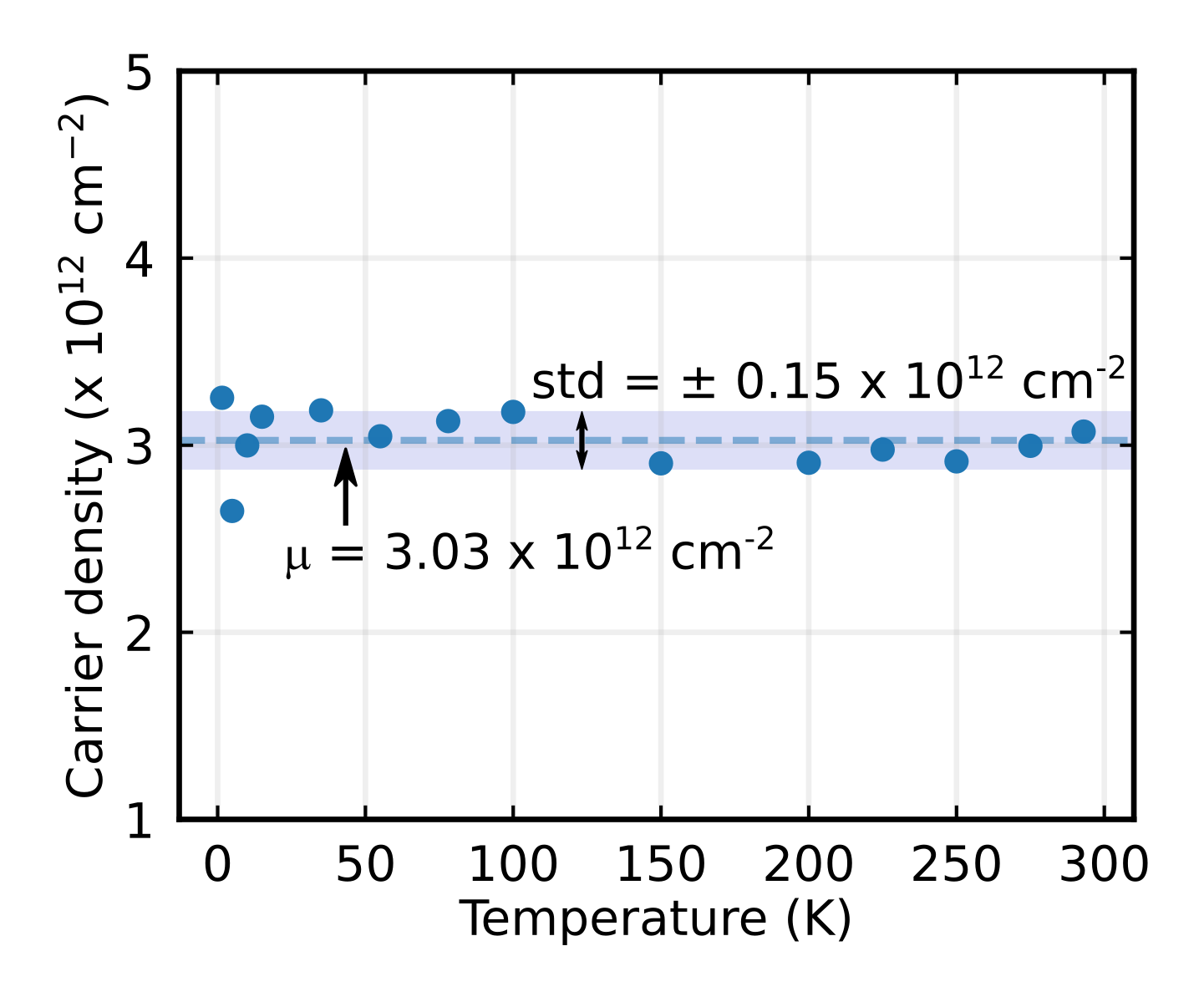}
		\caption[]{Carrier density measurement with respect to temperature.\\
			The carrier concentration is measured at $ V_g = 0 $~V with respect to temperature. Carrier concentration is measured to be temperature insensitive, which indicates that capacitance is also temperature insensitive according to the relation $ n = C(V_g - V_{dirac})/e $.}
		\label{sfig:carrier_temp}
	\end{figure}

	\begin{figure}[!htp]
		\centering			\includegraphics[width=144mm]{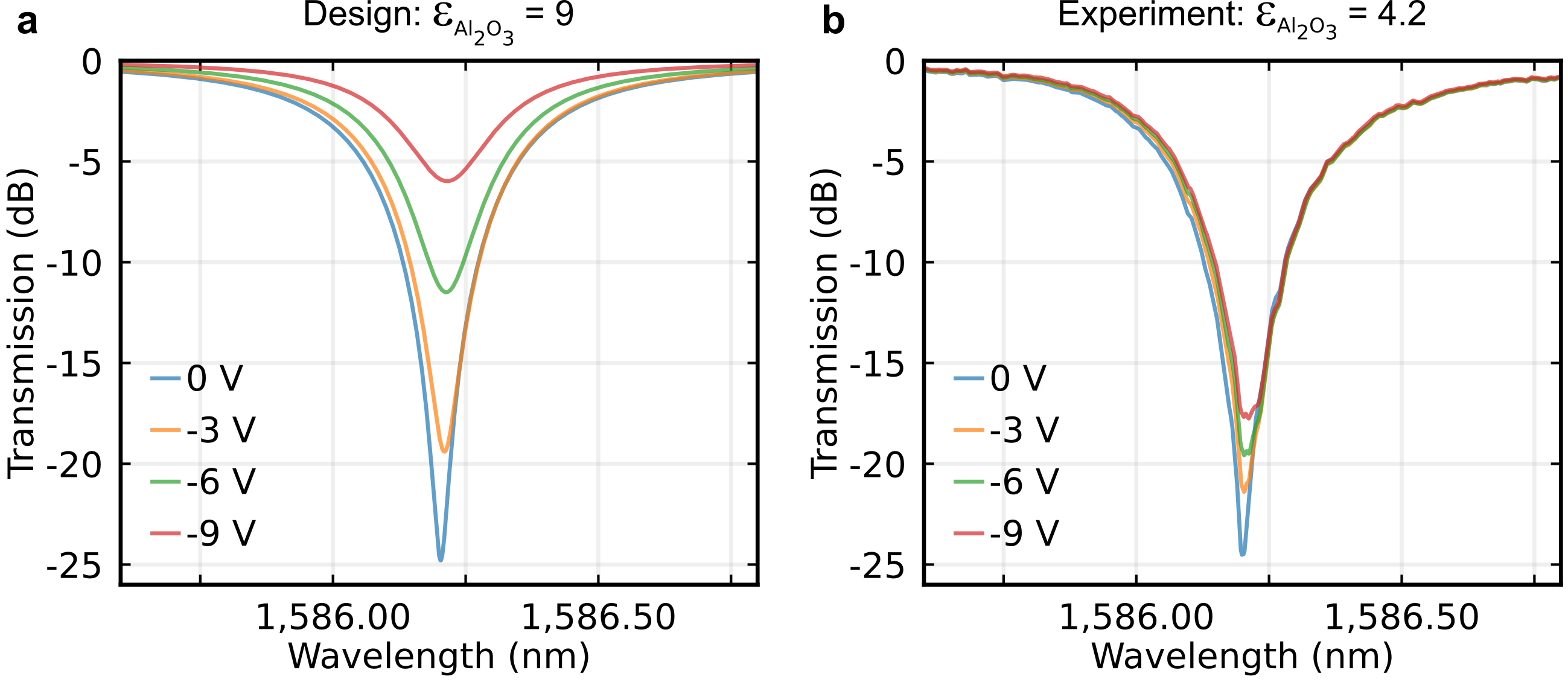}
		\caption[]{Transmission spectra of graphene modulator with different gate dielectric constant.\\
			(a) We design the graphene modulator with ALD \ch{Al2O3} gate dielectric constant of 9, typically reported in literature \citeSup{supyota2013characterization}. This enables modulation depth of around 19 dB over 9 V and insertion loss of less than 6 dB. (b) In our fabricated devices, we measure a dielectric constant of 4.2 (via Hall bar measurements) of our ALD \ch{Al2O3}. This reduced gate dielectric constant weakens the electrostatic gating and reducing control of graphene absorption, which leads to smaller modulation depth and higher insertion loss.} 
		\label{sfig:design}
	\end{figure}

	\bibliographystyleSup{unsrt}
	\bibliographySup{SReferences}
	
\end{document}